\documentclass{iopart}

\usepackage{cite}

\begin{document}

\title{Master equations for effective Hamiltonians}

\author{A B Klimov\dag, J L Romero\dag,
J Delgado\ddag and L L S\'{a}nchez-Soto\ddag}

\address{\dag\ Departamento de F\'{\i}sica,
Universidad de Guadalajara,
Revoluci\'on~1500, 44420~Guadalajara,
Jalisco, M\'{e}xico}

\address{\ddag\ Departamento de \'{O}ptica,
Facultad de F\'{\i}sica, Universidad Complutense,
28040~Madrid, Spain}

\eqnobysec

\begin{abstract}
We reelaborate on a general method for obtaining
effective Hamiltonians that describe different
nonlinear optical processes. The method exploits
the existence of a nonlinear deformation of the
su(2) algebra that arises as the dynamical symmetry
of the original model. When some physical parameter
(usually related to the dispersive limit) becomes
small, we immediately get a diagonal effective
Hamiltonian that represents correctly the dynamics
for arbitrary states and long times. We apply
the same technique to obtain how the noise terms
in the original model transform under this
scheme, providing a systematic way of including
damping effects in processes described in terms
of effective Hamiltonians.
\end{abstract}


\section{Introduction}

Effective Hamiltonians are employed frequently in
quantum optics. They appear most often in calculations
involving multiphoton processes~\cite{Shen67,Walls71}
or nonlinear optical effects~\cite{Yariv61,Tucker69}.
Notwithstanding that the authors who have taken recourse
to these Hamiltonians are truly innumerable, owing
to the great advantages they present in practice,
the procedure of their derivation has hardly been
given the attention which it merits. Even worse,
sometimes they are regarded as phenomenological
in nature and are not properly derived from the
underlying microscopic theory. For example, in
nonlinear optics it is customary to justify
effective Hamiltonians on the basis of a
quantized macroscopic theory of electrodynamics
in a nonlinear dielectric medium~\cite{Schenzle82},
which is not entirely satisfactory, especially
when one attempts to describe dispersive
media~\cite{Hillery84}.

Perhaps, the most common way of obtaining
effective Hamiltonians is via adiabatic
elimination~\cite{Cohen98,Walls95,Perina91}.
This approach was pioneered by Graham and
Haken~\cite{Haken68} for the parametric
oscillator with a medium consisting of
two-level atoms and the general strategy can
be stated as follows: starting from the exact
equations governing the interaction of a discrete
set of field modes with an ensemble of atoms,
one invokes the assumption that the atomic
polarization is a fast variable controlled by
the slow motion of the field amplitudes and
follows its evolution adiabatically (damping
is negligible in this time scale). Under
this hypothesis the atomic degrees of
freedom can be eliminated and the resulting
equations appear to be suffering a nonlinear
interaction that can be reinterpreted as the
Heisenberg equations of motion for the field
operator under the dynamics of an effective
nonlinear Hamiltonian.

This is a physically appealing picture,
but presents some serious drawbacks: first,
it does not provide a general prescription
for finding effective Hamiltonians since the
particular details strongly depend on the
model considered. Second, it could become
very cumbersome, and explicit but enormously
complicated expressions for the different
orders of approximation can be found in many
original publications~\cite{Bloembergen92,Shen85}.
Third, the procedure is not uniquely defined:
depending on the term eliminated, the outcome
of the final Hamiltonian could be
different~\cite{Puri88,Lugiato90,Negro99}.

For these compelling reasons, other methods
of deriving effective Hamiltonians exist.
For quantum optics, the ones devised
in~\cite{Sczaniecki83} and~\cite{Hillery85}
are especially germane. A thorough review
of alternative approaches may be found
in~\cite{Klein74}. Roughly speaking, all
the methods have in common that at some point
in their implementation one applies a unitary
(or canonical) transformation to the total
Hamiltonian and keeps only terms up to some
fixed order. Although sometimes the results
can be rather complicated, important examples
of application may be given~\cite{Shavitt80}.

Recently~\cite{Klimov00,Klimov02}, we have
provided a setting to support the theory of
effective Hamiltonians by resorting to some
elementary notions of group theory. The key
point for that is the simple observation that
most of the Hamiltonians in nonlinear quantum
optics contain cubic or higher terms in creation
and annihilation operators. Among others, typical
examples are $k$th harmonic generation, $k$-wave
mixing, and generalized Dicke models~\cite{Perina91}.
The common mathematical structure underlying
all these cases is a nonlinear or polynomial
deformation of su(2), which arises as the
dynamical symmetry algebra of the corresponding
Hamiltonian. This nonlinear algebra has recently
found an important place in quantum
optics~\cite{Karasiov92,Karasiov94,Debergh97,
Delgado00,Sunil00} because it allows us to handle
problems in very close analogy with the usual
treatment of an angular momentum. In particular,
we get a decomposition of the Hilbert space
into direct sums of invariant subspaces and
the dynamical problem generated by the corresponding
Hamiltonian can be reduced to the diagonalization
of a finite-dimensional matrix.

Unfortunately it is impossible to obtain, in general,
analytic expressions for the eigenvalues and
eigenstates of those matrices. Therefore, from
the dynamical viewpoint this algebraic structure
does not solve the problem. However, to some extent
it can help  to remedy it: if one applies a ``nonlinear
rotation" from this su(2) deformed algebra to the
original Hamiltonian, the transformed one is rather
involved, but when some physical parameter (dictated
by the model under consideration) becomes small, one
obtain an effective Hamiltonian that is diagonal and
represents correctly the dynamics for arbitrary
states and even for long times.

In this paper we wish to go one step further
showing a unique advantage of our method.
We stress that all the aforementioned
techniques implicitly assume that the
fields vary much more slowly than atomic
damping times. This means that the resulting
effective Hamiltonian can be used to describe only
short-time evolution; otherwise losses
begin to play an important role and cannot
be avoided~\cite{Gorbachev93,Gorbachev00}.

On the contrary, our method allows one to take
into account damping mechanisms in a natural way.
Indeed, starting from the original microscopic
model, we couple it to a reservoir and then
follow a standard procedure~\cite{Davies76}:
from the Liouville equation for the system
coupled to the reservoir, we trace over the
reservoir variables and, after a Markovian
approximation, we end up with a master equation.
This equation can be transformed by the same
nonlinear rotation as before, obtaining in
this way what we call an effective master
equation. In this paper we investigate this
method and present some relevant examples to
illustrate the procedure.

\section{Nonlinear rotations, effective
Hamiltonians and effective master equations}

To keep the discussion as self-contained
as possible and to introduce the physical
ideas underlying the method, let us start
with the simplest case of some physical
system whose Hamiltonian can be written as
\begin{equation}
\label{Hgen}
H = H_0 + H_{\mathrm{int}}
\end{equation}
where $H_0$ describes the free dynamics and
\begin{equation}
\label{Hint}
H_{\mathrm{int}} =
\Delta \ X_{3}+ g (X_+ + X_-) ,
\end{equation}
where $g$ is a coupling constant and $\Delta $
is a parameter usually representing the detuning
between frequencies of different subsystems,
although it is not necessary. The operators
$X_\pm$ and $X_3$ satisfy
\begin{equation}
\label{Xpm}
[ X_3 , X_\pm ] = \pm X_\pm ,
\qquad
[X_+, X_- ] = P(X_3),
\end{equation}
where $P(X_3)$ refers to an arbitrary polynomial
function of the diagonal operator $X_3$ with
coefficients perhaps depending on some integrals of
motion $N_j$. These commutation relations correspond
to the so-called polynomial deformation of su(2).
This kind of nonlinear algebras were discovered by
Higgs~\cite{Higgs79} and Sklyanin~\cite{Sklyanin82}
and have already played an important role in
several aspects of quantum mechanics~\cite{Rocek91,
Bonatsos93,Quesne95,Beckers96}.

Now suppose that for some physical reasons
(depending on the particular model under
consideration) the condition
\begin{equation}
\varepsilon =\frac{g}{\Delta }\ll 1
\end{equation}
is fulfilled. Then, it is clear that (\ref{Hint}) is
\textit{almost} diagonal in the basis that
diagonalizes $X_3$. In fact, a standard perturbation
analysis immediately shows that the first-order
corrections introduced by the nondiagonal part
$g(X_+ + X_-)$ to the eigenvalues of $X_{3}$
vanish and those of second order are proportional
to $\varepsilon \ll 1$. According to the technique
developed in \cite{Klimov00} , we apply the
following unitary transformation to equation (\ref{Hint})
\begin{equation}
\label{Umain}
U = \exp \left [ \varepsilon ( X_+ - X_- ) \right] ,
\end{equation}
which, in fact, is a small nonlinear rotation, in such
a way that
\begin{equation}
\label{H1}
H_{\mathrm{eff}} = U H_{\mathrm{int}}U^\dagger .
\end{equation}
Using the standard expansion $e^{A} B e^{-A} =
B + [A,B] + \frac{1}{2!} [A, [A,B]] + \ldots,$
and after keeping terms up to order $\varepsilon^2$,
we get
\begin{equation}
\label{H1eff}
H_{\mathrm{eff}} = \Delta \ X_3 +
\frac{g^2}{\Delta } P(X_3)  .
\end{equation}
The essential point is that we have got a
Hamiltonian that is diagonal in the basis
of eigenstates of $X_3$. In consequence,
the evolution (as well as the spectral)
problem is completely solved in this
approximation. Note that because (\ref{H1eff})
has the form of an expansion in the small
parameter $\varepsilon $, its eigenvalues will
coincide with those obtained using the standard
perturbation theory in the same order of
approximation. However, we stress that the
method is fully operatorial and avoids the
tedious work of computing the successive
corrections as sums over all the accessible
states.

To take into account damping mechanisms
in the original model, we couple it to
a reservoir and then follow a standard
procedure: from the Liouville equation for
the total system  we trace over the reservoir
variables and, after a Markovian approximation,
we end up with a master equation of the
form~\cite{Lindblad76,Orszag00}
\begin{equation}
\dot{\rho} = - i [H_{\mathrm{int}}, \rho] +
\sum_m \gamma_m \mathcal{L}[C_m]
\ \rho ,
\end{equation}
where $\gamma_m$ are real parameters determined
by the reservoir and $\mathcal{L}[C_m]$ is known
as the Lindblad superoperator
\begin{equation}
\mathcal{L}[C_m] \ \rho =
2 C_m \rho C_m^\dagger -
\{ C_m^\dagger C_m, \rho\} .
\end{equation}
Here $C_m$ are eigenoperators of the system
satisfying (in units $\hbar = 1$, which we shall
use throughout all this paper)
\begin{equation}
[H_0, C_m] = \omega_m C_m ,
\end{equation}
and their explicit form depends on the model
under consideration.

It seems natural to ask how this equation is
transformed by the same small rotation leading to
the effective Hamiltonian. Instead of discussing
an abstract formalism, we shall illustrate the
main ideas by resorting to some selected models
that are representative enough in quantum optics.

\section{Two-mode coupled oscillators}

The model of two coupled time-dependent harmonic
oscillators has been studied by many authors, who
have applied it to various problems of quantum
mechanics and quantum optics. For instance, it
was used to describe quantum amplifiers and
converters in~\cite{Gordon63,Mollow67,Lu73}.
The explicit exact solutions and propagators
of the Schr\"{o}dinger equation, as well as
solutions of the Heisenberg equations of motion,
were considered in~\cite{Abdalla87,Sandulescu87,
Yeon88,Abdalla90,Kim91}. Squeezing, photon
statistics, and entanglement in the system of
two coupled oscillators were studied in~\cite{Fan93,
Abdalla93,Manko94,Dantas95,Abdalla96,Kalmykov98}.
The model has been used recently to illustrate
the exchange of nonclassical properties~\cite{Dodonov01}
and the exchange of quantum states without energy
transfer between two modes of the electromagnetic
field~\cite{Oliveira99}.

The Hamiltonian for the model we are discussing
is
\begin{equation}
\label{Hco1}
H = \omega_a a^\dagger a + \omega_b b^\dagger b +
g ( a^\dagger b + b^\dagger a ) ,
\end{equation}
where $a (b)$ and $a^\dagger (b^\dagger)$ are
annihilation and creation operators for each
of one of the interacting modes of
frequency $\omega_a$ and $\omega_b$ and
$g$ is the coupling constant.

Because the excitation number
$N = a^\dagger a + b^\dagger b$
is an integral of motion we can
recast (\ref{Hco1}) in the form
(\ref{Hgen}) with
\begin{eqnarray}
\label{Hco2}
H_0 & = & \frac{1}{2}(\omega_a + \omega_b) N ,
\nonumber \\
& & \\
H_{\mathrm{int}} & = &
\frac{\Delta }{2} (b^\dagger b - a^\dagger a) +
g (a^\dagger b + b^\dagger a ) ,  \nonumber
\end{eqnarray}
where the detuning is $\Delta = \omega_b - \omega_a$.
But this is exactly of the form (\ref{Hint}) when
we introduce the operators
\begin{eqnarray}
& X_+ = b^\dagger a, \qquad
X_- = b a^\dagger  , &  \nonumber \\
& & \\
& X_3 = \frac{1}{2}
(b^\dagger b - a^\dagger a) , &  \nonumber
\end{eqnarray}
which coincide with the standard generators of the
su(2) algebra (without deformation).

Now suppose that one of the modes, say mode $b$,
is lossy, while mode $a$ is lossless. Coupling
the quantum oscillator $b$ to a standard reservoir
leads to the following master equation for the
density matrix in the interaction picture
\begin{equation}
\dot{\rho} = -i [ H_{\mathrm{int}}, \rho ] +
\frac{\gamma }{2} \mathcal{L}[b] \ \rho ,
\end{equation}
which has been thoroughly studied in the
literature~\cite{Cohen98,Walls95}.

From now on we shall be mainly interested
in the dispersive regimen, when
\begin{equation}
g \sqrt{( \bar{n}_a + 1 ) (\bar{n}_b+1)}
\ll |\Delta| ,
\end{equation}
where $\bar{n}_a$ and $\bar{n}_b$ are the
average number of excitations in each mode.
Then, we can apply the small rotation (\ref{Umain})
up to second-order terms obtaining
\begin{equation}
H_{\mathrm{eff}} = \Delta \ b^\dagger b +
\frac{g^2}{\Delta} ( b^\dagger b - a^\dagger a ).
\end{equation}
As stated before, our goal is to examine
how the master equation transforms under the same
small rotation. To this end, let us denote
\begin{equation}
\rho_{\mathrm{eff}} = U \rho U^\dagger .
\end{equation}
Then, after some calculations, one gets that
the new master equation becomes
\begin{eqnarray}
\dot{\rho}_{\mathrm{eff}} & = &
-i [ H_{\mathrm{eff}}, \rho_{\mathrm{eff}} ]
\nonumber \\
& + &
\frac{\gamma }{2} \left( 1 -
\frac{g^2}{2 \Delta^2} \right) \mathcal{L}[b] \
\rho_{\mathrm{eff}} +
\frac{\gamma }{2}\frac{g^2}{\Delta^2} \mathcal{L}[a] \
\rho_{\mathrm{eff}} .
\end{eqnarray}g5gg
We can see here an effective transfer of
decoherence for the lossy mode $b$ to mode $a$,
which is of order $g^2/\Delta^2$. Note that second-order
corrections to $H_{\mathrm{eff}}$ are of the same
order of magnitude, but they play no significant
role because are diagonal.

If we further assume that mode $b$ is initially
in vacuum we have no effective contributions from
the normally-ordered terms like $b^\dagger b$ and
the effective Hamiltonian reduces to
\begin{equation}
H_{\mathrm{eff}} = - \frac{g^2}{\Delta} \ a^\dagger a,
\end{equation}
while the master equation reduces to
\begin{equation}
\dot{\rho}_{\mathrm{eff}} = -i
[ H_{\mathrm{eff}}, \rho_{\mathrm{eff}} ] +
\frac{\gamma }{2}\frac{g^2}{\Delta^2}
\mathcal{L}[a] \ \rho_{\mathrm{eff}} .
\end{equation}
The transfer of decoherence between modes is
here even more striking. Some efforts,
especially with the method of stochastic
unravellings~\cite{Plenio98}, have been made
to elucidate the physical contents of this
kind of equations. We stress, however, that
is not the scope of this paper to work out
their solutions, but rather to show how
they appear in the context of effective
Hamtiltonians.

\section{Second-harmonic generation}

Second-harmonic generation is the simplest
nonlinear optical process. It exhibits a rich
spectrum of nonclassical features such as
photon antibunching, squeezing, or collapses
and revivals~\cite{Kozierovski77, Mandel82,
Wu86,Drobny92,Drobny93}. Nikitin and
Masalov~\cite{Nikitin91} have shown that,
at resonance, the quantum state of the
fundamental mode evolves into a superposition
of two macroscopically distinguishable states;
i.e., a Schr\"{o}dinger cat state. This point
may be considered unrealistic, because the
stringent experimental constraints of having
perfect phase matching and low decoherence could
be rather difficult to attain.

In \cite{Klimov01} we  have considered the
dispersive limit of second-harmonic generation,
which seems to be almost ignored in the literature.
Without assuming perfect resonance, this process
is described by the following model
Hamiltonian
\begin{equation}
\label{2hg1}
H = \omega_a a^\dagger a + \omega_b b^\dagger b
+ g ( a^2 b^\dagger + a^\dagger {}^2 b ) .
\end{equation}
The excitation number $N= a^\dagger a + 2 b^\dagger b$,
is also a constant of motion. If the detuning is taken
now as $\Delta = \omega_b - 2 \omega_a$, the Hamiltonian
(\ref{2hg1}) can be rewritten in the general form
(\ref{Hgen}) with
\begin{eqnarray}
\label{2hg2}
H_0 & = & \frac{1}{3} (\omega_b + \omega_a) N ,  \nonumber \\
& & \\
H_{\mathrm{int}} & = & \frac{\Delta }{3}
(b^\dagger b - a^\dagger a ) +
g ( a^2 b^\dagger + a^\dagger {}^2 b ) .  \nonumber
\end{eqnarray}
The polynomial deformation of su(2) naturally emerges
when we rewrite
\begin{eqnarray}
& X_+ = b^\dagger a^2 ,
\qquad
X_- = a^\dagger {}^2 b , &
\nonumber \\
& & \\
& X_3 = \frac{1}{3}
(b^\dagger b - a^\dagger a ) . &  \nonumber
\end{eqnarray}

Let us next assume that we are in the dispersive
limit
\begin{equation}
g( \bar{n}_a + 1 ) \sqrt{\bar{n}_b+1}
\ll | \Delta | ,
\end{equation}
and apply the small nonlinear rotation (\ref{Umain}).
The final result is~\cite{Klimov01}
\begin{equation}
\label{HeffK}
H_{\mathrm{eff}} =
\frac{\Delta}{3} (b^\dagger b - a^\dagger a ) +
\frac{g^2}{\Delta}
[4 b^\dagger b a^\dagger a - (a^\dagger a)^2 ] .
\end{equation}
This Hamiltonian is diagonal, which implies that
there is no population transfer between the modes,
as it would be expected in the far-off resonant
limit. The first term does not affect the dynamics
and just leads to a rapid oscillation of the wave
function.

When the harmonic mode $b$ is initially in the
vacuum, the two terms containing $b^\dagger b$
do not contribute. In addition, the linear term
$a^\dagger a$ leads just to a $c$-number phase
shift and can be also omitted. In consequence,
(\ref{HeffK}) reduces to
\begin{equation}
\label{ide}
H_{\mathrm{eff}} = - \frac{g^2}{\Delta}
(a^\dagger a)^2 ,
\end{equation}
which is nothing but the interaction Hamiltonian
that governs the state evolution of the single-mode
field $a$ in a Kerr medium. We emphasize that Kerr
effect provides a nonlinearity of particular interest
for generating field cat states~\cite{Kitagawa86,
Milburn86,Tanas91,Tara93,Varada93}. Apart from
their intrinsic simplicity, Kerr-based schemes
have the specific advantage of not relying on
conditional measurements. However, the realistic
values of Kerr coefficients are quite small,
thus requiring a large interaction length. Then,
losses become significant and may destroy the
very delicate quantum superpositions. In short,
although very appealing from the physical viewpoint,
Kerr schemes are not generally considered to be
realistic.

For this reason, the identification proposed
in (\ref{ide}) could be more than an academic
curiosity: second-harmonic generation is, for
a variety of reasons~\cite{White00}, more robust
than the Kerr effect as for noise-limiting factors.
In consequence, this scheme could be an experimentally
feasible proposal to generate optical cat states.
However, it is clear that losses should be
taken into account. While this is standard in a
Kerr medium, for the case of the dispersive
limit of second-harmonic generation is far from
being known. In our approach, by transforming the
standard master equation  (under the same condition
of harmonic mode $b$ initially in the vacuum)
one easily gets that
\begin{equation}
\dot{\rho}_{\mathrm{eff}} = - i
[ H_{\mathrm{eff}}, \rho_{\mathrm{eff}} ]
+ \frac{\gamma}{2} \frac{g^2}{\Delta^2}
\mathcal{L}[a^2] \ \rho_{\mathrm{eff}} ,
\end{equation}
which, in fact, agrees with the standard form of
dealing with decoherence effects in a Kerr medium,
in which all the dissipation (up to order $g^2/\Delta^2$)
takes place via two-photon process.

\section{Dicke model}

Finally, let us consider the example of the
well-known Dicke model describing the interaction
of a single-mode field of frequency $\omega_{\mathrm{f}}$
with a collection of $A$ identical two-level atoms
with transition frequency $\omega_0$. Making the
standard dipole and rotating-wave approximations,
the model Hamiltonian reads as
\begin{equation}
\label{Dicke1}
H = \omega_{\mathrm{f}}a^\dagger a +
\omega_0 S_3 +
g (a S_+ + a^\dagger S_- ),
\end{equation}
where $(S_\pm ,S_3)$ are collective atomic
operators forming an $(A+1)$-dimensional
representation of su(2).

Now the excitation number reads as $N = a^\dagger a
+ S_3$, and once again we can recast (\ref{Dicke1})
in the form (\ref{Hgen}) with
\begin{eqnarray}
\label{HD}
H_0  & = & \omega_{\mathrm{f}} N ,
\nonumber \\
& & \\
H_{\mathrm{int}} & = &
\Delta \ S_3 +
g ( aS_+ + a^\dagger S_- ) ,  \nonumber
\end{eqnarray}
the detuning being $\Delta = \omega_0 -
\omega_{\mathrm{f}}$. The operators
defining the su(2) deformation are given by
\begin{eqnarray}
& X_+ = a S_+ , \qquad X_- = a^\dagger S_- , &  \nonumber \\
& & \\
& X_3 = S_3 , &  \nonumber
\end{eqnarray}
and then the interaction part of the Dicke
Hamiltonian looks exactly as in (\ref{Hint}).

To keep our discussion as realistic as possible,
let us consider the case in which the atomic
system and the field interact in a lossy cavity,
in such a way that photons leakage occurs say
through a partially transmitting mirror with
a decay rate given by $\gamma$. In the case of
optical frequencies, thermal excitation from
the environment of the continuum of modes outside
the cavity is negligible and the dynamics is well
described by the master equation~\cite{Gardiner00}
\begin{equation}
\label{meqor}
\dot{\rho} = -i [ H_{\mathrm{int}}, \rho ]
+ \frac{\gamma }{2} \mathcal{L}[a] \ \rho .
\end{equation}
This decay is also responsible for the
rapid decay of any eventual quantum coherence
generated within the cavity.

The dispersive limit may be stated now in
the form~\cite{Brune96}
\begin{equation}
A g \sqrt{\bar{n}+1} \ll | \Delta | ,
\end{equation}
where $\bar{n}$ is the average number of photons
in the field. Then, the nonlinear rotation
(\ref{Umain}) transforms the Hamiltonian~(\ref{HD})
into~\cite{Klimov00}
\begin{equation}
\label{H2}
H_{\mathrm{eff}} = \Delta \ S_3 +
\frac{g^2}{\Delta} [ S_3^2 - 2 (a^\dagger
a +1 ) S_3 - C_2 ] ,
\end{equation}
where $C_2 = A/2 ( A/2+1) $ is the value of the
Casimir operator for su(2). The Hamiltonian (\ref{H2})
was previously obtained in \cite{Agarwal97} by quite
a different method (see also \cite{Klimov98}) and,
due to the presence of the nonlinear term $S_3^2$,
has been considered as a candidate for the generation
of atomic Schr\"{o}dinger cats and squeezed
states~\cite{Ueda93}.

Transforming the master equation (\ref{meqor}) by
the operator (\ref{Umain}) we obtain
\begin{equation}
\dot{\rho}_{\mathrm{eff}} = - i
[H_{\mathrm{eff}}, \rho_{\mathrm{eff}}] +
\mathcal{L}[a] \ \rho _{\mathrm{eff}} +
\varepsilon \mathcal{L}_1[a]\
\rho _{\mathrm{eff}}  +
\varepsilon^2 \mathcal{L}_2[a] \
\rho _{\mathrm{eff}},
\end{equation}
where
\begin{eqnarray}
\mathcal{L}_1 [a] \ \rho_{\mathrm{eff}}  & = &
2(S_- \rho_{\mathrm{eff}} a^\dagger +
a \rho_{\mathrm{eff}} S_+ )  \nonumber \\
& - & a^\dagger  S_- \rho _{\mathrm{eff}} -
S_+ a \rho_{\mathrm{eff}} - \rho _{\mathrm{eff}}
a^\dagger S_- - \rho _{\mathrm{eff}} a S_+ ,
\nonumber \\
\mathcal{L}_2[a] \ \rho _{\mathrm{eff}} & = &
2 S_- \rho_{\mathrm{eff}} S_+ -
S_+ S_- \rho_{\mathrm{eff}} -
\rho_{\mathrm{eff}} S_+ S_-
\nonumber \\
& + & 2 S_3 a \rho_{\mathrm{eff}}
a^\dagger + 2 a \rho_{\mathrm{eff}} a^\dagger
S_3 - 2 a^\dagger  a S_3 \rho_{\mathrm{eff}} -
2 \rho_{\mathrm{eff}} S_3 a^\dagger a.
\end{eqnarray}
In the rotating frame, the term $\mathcal{L}_1$
oscillates very rapidly and consequently can be
eliminated by the corresponding transformation
to a rotating frame of $\rho _{\mathrm{eff}}$.
Surprisingly, the term $\mathcal{L}_2$ contains
a resonant part that is time independent in
the rotating frame. This term generates effective
atomic dissipation, leading to a complete
decoherence for times or the order of
$1/(\varepsilon ^{2}\gamma ).$

To further proceed, let us assume that initially
the field is in vacuum, while the atom is in an
arbitrary atomic state $\rho _{\mathrm{at}}(0)$.
One could expect that in this dispersive limit,
because there is no net energy transfer between
the atoms and the field, the atomic system would
not feel the field. One can check that
\begin{equation}
\dot{\rho}_{\mathrm{eff}}= - i
[H_{\mathrm{eff}},\rho_{\mathrm{eff}} ] +
2 \frac{\varepsilon^{2}}{\gamma }
(S_- \rho_{\mathrm{eff}} S_+ -
S_+ S_- \rho_{\mathrm{eff}} -
\rho_{\mathrm{eff}} S_+ S_- ) ,
\end{equation}
which shows the appearance of an effective
atomic dissipation due to the low-amplitude
atom-field transitions generated by corrections
to the initial state due to the transformation
by the nonlinear rotation $U$.

\section{Concluding remarks}

In this paper we have studied a systematic way of
including damping in processes described by effective
Hamiltonians. Against the widespread opinion that
noise cannot be included in this kind of theories
(and, therefore, they describe only short-time evolution)
we have shown how to obtain an effective master
equation of the Lindblad type that allows one to
take into account  the dissipation from first
principles and not merely adding phenomenological
terms.

Here we have considered three specific models of
interest in quantum optics, altough it is easy to
convince oneself that our results can be generalized
to any system that can be described in terms of
a polynomial deformation of the algebra su(2).
We finally stress that we have have not discussed
in detail the effects of this dissipation for each
model under consideration, since our main goal
has been only to provide the adequate framework to
deal with such effects.

\end{document}